COOL DWARFS IN WIDE MULTIPLE SYSTEMS

# PAPER 2: A DISTANT M8.5V COMPANION TO HD 212168 AB

*By José A. Caballero*
*Centro de Astrobiología (CSIC-INTA), Madrid*
*and David Montes*
*Universidad Complutense de Madrid, Madrid*

DENIS J222644.3–750342 is a nearby, very-low-mass, M8.5 V-type star that has been repeatedly targeted by kinematic and activity studies. Thought to be an isolated star, it is actually a wide common-proper-companion at 265 arcsec from the bright G0 V *Hipparcos* star HD 212168. The third member in the trio is CD–75 1242, a poorly investigated K dwarf. We confirm the physical binding of the triple system, to which we call Koenigstuhl 5, by compiling common radial velocities and proper-motions and measuring constant angular separations and position angles between components A, B and C over very long time baselines (A-B: 114 a; A-C: 22 a). With about 0.095 $M_{sol}$, the M8.5 V star at 6090 AU to the G0 V primary is one of the least-bound, ultracool dwarfs in multiple systems.

## *Koenigstuhl 5 ABC: two Sun-like* Hipparcos *stars with an M8.5V companion*

We continue the series of works devoted to investigating cool dwarfs in wide multiple systems started by Caballero[1]. In this issue we present a new system containing a pair of relatively bright Sun-like stars and a very-low-mass M8.5V star 4.4 arcmin southeast. It was found during a kinematic investigation of ultracool dwarfs in young moving groups[2], but has not been formally reported yet.

The central pair, formed by the stars HD 212168 and CD–75 1242, was discovered by Dunlop[3] in 1826 and received the *Washington Double Star Catalogue*[a] (WDS[4]) code and number Dun 238. Although the *WDS* remarks that "proper motion or other technique indicates that this pair is physical" and Raghavan *et al.*[5] added that the two stars in the system "[match] proper motion and photometric distance", there has been no consensus on the actual binarity of the pair. Sinachopoulos[6] found that the B component has Strömgren *uvby* colours that were much too red for a its spectral type, thought to be "G0– V" (with a slash), from where Gray *et al.*[7] concluded that "since 'A' is also a G0 dwarf and is about 2.7 mag brighter, this is undoubtedly an optical pair".

In Table I, we compile the basic astrophysical data of HD 212168 and CD–75 1242 from a number of sources (HIP2: *Hipparcos*-2[8]; TYC: Tycho-2[9]; USNO-B1.0: US Naval Observatory-B[10]; 2MASS: Two-Micron All-Sky Survey[11]; Ev64[12]; Ni78[13]; Ob78[14]; He96[15]; No04[16]; Gr06[7]; Je06[17]; Ho09[18]; Ca11[19]). While the observed

---

[a] http://ad.usno.navy.mil/wds/

parameters match A being an inactive, early G-type star, non-Strömgren (*i.e.*, Johnson, Geneva) photometry also suggests a later spectral type for B, which we estimate to be an early or intermediate K dwarf (or, more accurately, K3-4: V, where the colon stands for uncertain spectral value). The spectra of both stars available at the NStar Spectra Homepage[b] are basically the same, except for what seems to be a normalization problem redwards of 470 nm. Gray *et al.*[7] probably made a book-keeping slip in observing the star: since the two stars have *Hipparcos* numbers several units apart, they could have observed them on separate occasions, and mistakenly observed the primary twice, obtaining essentially the identical spectral type from both spectra. CD–75 1242 was classified as type G5 in the *Cape Photographic Catalogue*, where non-Henry Draper types were provided by Margaret Mayall from Harvard objective-prism plates. This usually implies a modern Morgan-Keenan type of K0, and the overlapping spectra probably led to the type being estimated too early. Given the brightness of CD–75 1242, of $V$ = 8.7 mag, about 2.7 mag fainter than HD 212168, it should be easy to obtain a new mid-resolution spectrum from the Southern Hemisphere to solve this dilemma.

The primary, HD 212168, has not low-mass, close companions based on deep adaptive-optics imaging with NACO[20] and on accurate, long-term radial-velocity monitoring[21], nor flux excess in the mid-infrared[22].

The *Hipparcos*-based parallax of CD–75 1242 is relatively poor because of interference from the companion (a common problem in the *Hipparcos* catalogue). However, both HD 212168 and CD–75 1242 have the same radial velocity within uncertainties. But the definitive confirmation that they form "undoubtedly a *physical* pair" comes from a simple astrometric analysis consisting in measuring the angular separation between the two components with a very long time baseline. We compile all available astrometric epochs in Table II. We measured angles from HIP2[8], TYC[9], 2MASS[11] and AC2000.2[23] and took observed angles from the original van Albada-van Dien works (vAd1983[24], vAd1985[25], vAd1987[26]). The remaining epochs were either tabulated by the *WDS* catalogue[c] (according to WDS nomenclature: Dun1829[3], HJ_1847[27], WFC1998[23], ByS1900[28], Daw1918[29], Ged1940g[30], WFC1966c[31], WFC1971[32], The1975[33], WFC1992[34]) or kindly provided by the anonymous referee (Gou1886[35], *Akari*[36]). For some unknown reason, two Deep Near-Infrared Southern Sky Survey (DENIS[37]) astrometric epochs in 1999.81 smeared the results and were not used.

A total of 114.2 a (over a century) elapsed between the first and last reliable astrometric epochs in Table II (1892.78 and 2007.0). If only the most accurate measurements after 1945 are taken into account, the mean angular separation and position angle turn to be $<\rho>$ = 20.56±0.11 arcsec, $<\theta>$ = 80.2±0.5 deg, showing CD–75 1242 to be fixed relative to the primary. The large scatter in position angle is because some authors (*e.g.*, van Albada-van Dien[24-26]) reported measurements at-epoch (*i.e.*, not precessed to a common equinox), rather than for J2000, which is the case for the other measuremens shown. Besides, small time variations of the order of what is observed in $\rho$ and $\theta$ are expected from two Sun-like stars separated by slightly less than 500 AU. Our astrometric follow-up covered approximately 1.5% of a seven-millenia orbit.

---

[b] `http://stellar.phys.appstate.edu/`

[c] This research has made use of the Washington Double Star Catalog maintained at the U.S. Naval Observatory.

The third component in the system, DENIS J222644.3–750342, was discovered and first investigated by Phan-Bao *et al.*[38,39]. The star turned to be an M8.5 dwarf with faint H$\alpha$ emission[40,41,42,43], relatively large rotational velocity[42] and thought to be single and isolated from any other star. See Table III for a compilation of astrophysical parameters of the ultracool dwarf.

While the radial velocity of DENIS J222644.3–750342 measured by Reiners & Basri[44] matches within 1$\sigma$ with the one measured for HD 212168, reported heliocentric distances are significantly shorter than for the G0 V star, for which there is a parallax measurement by *Hipparcos*. In particular, different distances in the interval from 15.9 to 19.4 pc have been estimated from spectral type-magnitude relations by virtually all authors that have investigated the ultracool dwarf[39-43,45,46], with most estimations around 17 pc. Being a companion to HD 212168 AB translates into being located significantly further, at 23.0±0.3 pc.

Does DENIS J222644.3–750342 have the same proper motion as HD 212168? First, we have to solve a mess related to proper-motions of the three involved stars, as illustrated by Table IV:

- As seen above, HD 212168 and CD–75 1242 have (roughly) the same proper-motion, in spite of the large proper-motion uncertainties in the re-reduced *Hipparcos*[8] data and, especially, in Tycho-2[9], USNO-B1.0[10] and the original *Hipparcos* data[47], which we do not list to avoid confusion.
- The reference for the proper-motion of DENIS J222644.3–750342 that Faherty *et al.*[46] used for their Brown Dwarf Kinematics Project is the NLTT Catalogue[48], but they actually took it from Phan-Bao *et al.*[39].
- *Simbad* lists only the proper motion of DENIS J222644.3–750342 from Schmidt *et al.*[41], which has very large uncertainties and is quite different from the one from Phan-Bao *et al.*[39].
- The proper-motion from Phan-Bao *et al.*[39] resembles the ones from USNO-B1.0 and Positions and Proper-motions Extended Large (PPMXL[49]), which in turn look like the re-reduced *Hipparcos* proper motion of HD 212168.

We used 2MASS, DENIS and the SuperCOSMOS[50] digitizations of the photographic plates from the *United Kingdom Schmidt Telescope* (UKST) and European Southern Observatory red plate (ESO Red) as in Caballero[51] for measuring a new accurate proper motion of DENIS J222644.3–750342. In particular, we used seven astrometric epochs spaced by 22.15 a (the ones listed in Table V plus another DENIS epoch on 1999 Aug 28). In contrast, previous authors had used only part of our dataset (e.g., Schmidt *et al.*[41] used *UKST* plates of the Digitized Sky Survey and 2MASS). To sum up, the USNO-B1.0 and PPMXL measurements and ours, (+53.4 ± 0.9, +6.5 ± 0.7) mas a$^{-1}$, are consistent with HD 212168 AB and DENIS J222644.3–750342 having a common proper-motion (Table IV).

To confirm our assumption, we studied the constancy of the angular separation and position angle between HD 212168 and DENIS J222644.3–75034. As shown in Table V, $\rho$ and $\theta$ are indeed constant with root-mean-squares of only 0.10 arcsec and 0.05 deg over six astrometric epochs spaced 22.15 a. The proper motion of the system is large enough that there would be significant change in $\rho$ and $\theta$ over the 22-year baseline of the observations if the two stars were not in fact linked. At this point, we conclude already that the three stars, HD 212168 (G0 V), CD–75 1242 (K3-4: V) and DENIS J222644.3–750342 (M8.5 V) form a triple system that we called Koenigstuhl 5 AB-C (KO 5), following the nomenclature introduced by Caballero[1,52].

In Tables I and III we provide masses of CD–75 1242 and Koenigstuhl 5 C derived from photometry and theoretical models[53,54] assuming that they are located at the same distance as HD 212168 and have a Sun-like age and metallicity. Although the primary *UVW* Galactic space velocities derived by Holmberg *et al.*[18] are consistent with membership in the Ursa Major moving group ($\tau \sim$ 0.3 Ga)[55] and in the young disc ($\tau \leq$ 1 Ga), no features of youth have been identified up to now in any of the three components. See also Table I for the age estimated for the primary by Casagrande *et al.*[19]. Because of the dwarf class of the primary, its solar metallicity ([Fe/H] = –0.05) and low vertical Galactic velocity component ($W$ = –12 km $s^{-1}$), the system cannot be extremely old (i.e., $\tau$ > 10 Ga).

At $d$ = 23.0±0.3 pc, Koenigstuhl 5 C seems to be overluminous with respect to the spectral type-magnitude relations used previously. The star appears single at the resolution of the 2MASS images, where an equal pair even ~1 arcsec separation would be obvious. This does not rule out a closer binary.

Koenigstuhl 5 C is located at a projected physical separation to the primary of over 6000 AU. This separation is not extraordinary (see, e.g., recent works by Caballero[56,57] and Shaya & Olling[58]), but it is considerable for an ultracool dwarf of 0.095±0.05 $M_{sol}$ close to the substellar boundary (Table VI). There are few comparable systems, being perhaps V1054 Oph ABC + GJ 643 + vB 8 (a group of five M dwarfs including an M7 V star at 1500 AU to the triple primary[59]), η CrB AB-C (an L8 V brown dwarf at 3600 AU of a G1 V+G3 V pair[60]) and Koenigstuhl 3 A-BC (an M8 V+L3 V pair at 11900 AU to an F8 V[61]) the most representative ones. Koenigstuhl 5 AB-C, with an approximate binding energy of –54 $10^{33}$ J if the three components are taken into account, is one of the least bound systems known (see, *e.g.*, Fig. 2 in Caballero[56]). Our triple system goes on with the Koenigstuhl series of wide multiple systems with ultracool dwarfs that represent a challenge for formation scenarios and stability models of fragile systems as they travel across the Galactic disc. Besides, the hypothetical overluminosity of Koenigstuhl 5 C could be explained by unresolved equal binarity and, thus, our system would be a hierarchical quadruple system with roughly twice the estimated binding energy. Dedicated high-resolution imaging of the system components would be of great interest.

Table I

*Basic data of components A and B*

| *Datum* | *A* | *B* | *Origin* |
|---|---|---|---|
| Name | HD 212168 | CD−75 1242 | Simbad |
| HIP | 110712 | 110719 | HIP2 |
| $\alpha$ (J2000) | 22 25 51.16 | 22 25 56.40 | HIP2 |
| $\delta$ (J2000) | −75 00 56.5 | −75 00 52.8 | HIP2 |
| $d$ [pc] | 23.0±0.3 | 18±2 (23.0±0.3) | HIP2 |
| $U$ [mag] | 6.82 | ... | Ni78 |
| $B_T$ [mag] | 6.839±0.015 | 10.134±0.027 | TYC |
| $B$ [mag] | 6.68 | 9.82 | Ni78, *this work* |
| $V_T$ [mag] | 6.187±0.010 | 8.845±0.015 | TYC |
| $V$ [mag] | 6.04 | 8.73 | Ni78, *this work* |
| $R_F$ [mag] | 5.78 | 8.10 | USNO-B1.0 |
| $I_N$ [mag] | 5.48 | 7.55 | USNO-B1.0 |
| $J$ [mag] | 5.262±0.276 | 6.559±0.029 | 2MASS |
| $H$ [mag] | 4.768±0.021 | 5.937±0.021 | 2MASS |
| $K_s$ [mag] | 4.705±0.016 | 5.809±0.023 | 2MASS |
| $b-y$ [mag] | 0.374 | 0.674 | Ob78 |
| $m_1$ [mag] | 0.197 | 0.555 | Ob78 |
| $c_1$ [mag] | 0.357 | 0.116 | Ob78 |
| $\beta$ [mag] | 2.614 | 2.520 | Ob78 |
| $v_r$ [km s$^{-1}$] | +13.2±0.7 | +13±2 | No04, Ev64 |
| $U,V,W$ [km s$^{-1}$] | +2, −8, −12 | (+2, −8, −12) | Ho09 |
| $T_{eff}$ | 5940±80 | ... | Ca11 |
| log$g$ | 4.31 | ... | Ca11 |
| [Fe/H] | −0.05 | ... | Ca11 |
| log $R'_{HK}$ | −4.89 to −4.97 | ... | He96, Je06 |
| Sp. Type | G0 V | K2-3: V | Gr06, *this work* |
| $\tau$ (Ga) | $6^{+2}_{-3}$ | 1 to 10 | Ca11, *this work* |
| $M_V$ | 4.22±0.03 | 6.92±0.03 | *This work* |
| $M$ [M$_{sol}$] | 1.10±0.10 | 0.80±0.10 | *This work* |

Table II

*Astrometric observations of the AB pair (Dun 238 AB)*

| *Epoch* | *ρ [arcsec]* | *θ [deg]* | *Origin* |
|---|---|---|---|
| 1826 | 14 | 90 | Dun1829 |
| 1835.73 | 25 | 83.9 | HJ_1847a |
| 1836.24 | 18.09 | 82.9 | HJ_1847b |
| 1882.785 | 19.2 | 83 | Gou1886 |
| 1892.78 | 20.134 | 82.6 | WFC1998 |
| 1893.707 | 19.9±0.5 | 80.4±0.2 | AC2000.2 (WFC1998) |
| 1894.73 | 20 | 85 | ByS1900 |
| 1894.74 | 19.713 | 79.9 | WFC1998 |
| 1917.92 | 20.12 | 81.2 | Daw1918 |
| 1940.60 | 20.2 | 79.8 | Ged1940g |
| 1947.80 | 20.532 | 80.8 | WFC1966c |
| 1956.19 | 20.501 | 81.4 | WFC1971 |
| 1958.567 | 20.419 | 80.25 | The1975 |
| 1971.49 | 20.478 | 80.1 | WFC1992 |
| 1975.514 | 20.48±0.01 | 80.15±0.04 | vAd83 |
| 1976.462 | 20.451±0.017 | 80.30±0.07 | vAd85 |
| 1983.640 | 20.695±0.007 | 79.70 | vAd87 |
| 1991.250 | 20.6±0.4 | 79.6±0.2 | HIP2 |
| 1991.500 | 20.59 | 79.8 | TYC |
| 1999.930 | 20.75±0.13 | 79.6±0.2 | 2MASS |
| 2007.0 | 20.71 | 80.3 | *Akari* |

Table III

*Basic data of component C*

| *Datum* | *C* | *Origin* |
|---|---|---|
| Name | DENIS J2226443–750342, Koenigstuhl 5 C | P-B03, *this work* |
| α (J2000) | 22 26 44.41 | 2MASS |
| δ (J2000) | –75 03 42.5 | 2MASS |
| *d* [pc] | 15.9 to 19.4 (23.0±0.3) | Various authors |
| $\langle R_F \rangle$ [mag] | 18.7 | USNO-B1.0 |
| $\langle i \rangle$ [mag] | 15.21 | DENIS |
| $I_N$ [mag] | 15.05 | USNO-B1.0 |
| *J* [mag] | 12.353±0.023 | 2MASS |
| *H* [mag] | 11.696±0.027 | 2MASS |
| $K_s$ [mag] | 11.246±0.023 | 2MASS |
| $v_r$ [km s$^{-1}$] | +14.7±3.0 | RB09 |
| $M_J$ | 10.54±0.03 | *This work* |
| Sp. type | M8.5 V | Various authors |
| *pEW*(Ha) | −0.6 to −7.1 | Various authors |
| *M* [$M_{sol}$] | 0.095±0.005 | *This work* |

Table IV

*Proper motions of components A, B and C*

| *Component* | $\mu_\alpha \cos\delta$ *[mas a$^{-1}$]* | $\mu_\delta$ *[mas a$^{-1}$]* | *Reference* |
|---|---|---|---|
| A | +57.79±0.50 | +12.25±0.39 | HIP2 |
| B | +67.78±5.59 | +9.09±4.82 | HIP2 |
| C | +46±10 | +22±10 | USNO-B1.0 |
| | +60.6±9.3 | +16.6±9.3 | PPMXL |
| | +48±19 | +14±19 | P-B03 |
| | +20±55 | +46±24 | Sc07 |
| | +53.4±0.9 | +6.5±0.7 | *This work* |

Table V

*Astrometric observations of the AC pair (KO 5 AC)*

| *Epoch* | *ρ [arcsec]* | *θ [deg]* | *Origin* |
|---|---|---|---|
| 1977 Oct. 13 | 264.9±0.2 | 128.8±0.2 | UKST Blue |
| 1984 Oct. 12 | 264.8±0.2 | 128.9±0.2 | ESO Red |
| 1993 Sep. 25 | 264.8±0.2 | 128.8±0.2 | UKST Infrared |
| 1996 Oct. 10 | 264.7±0.2 | 128.9±0.2 | UKST Red |
| 1999 Oct. 24 | 264.9±0.1 | 128.9±0.2 | DENIS |
| 1999 Dec. 07 | 264.71±0.13 | 128.8±0.2 | 2MASS |

Table VI

*Basic data of the AB and AC pairs*

| *Pair* | *A–B (Dun 238 AB)* | *A–C (KO 5 AC)* |
|---|---|---|
| <*ρ*> [arcsec] | 20.56±0.11 | 264.79±0.10 |
| <*θ*> [deg] | 80.2±0.5 | 128.84±0.05 |
| *d* [pc] | 23.0±0.3 | |
| *s* [AU] | 476±6 | 6090±80 |
| $M_{total}$ [$M_{sol}$] | 1.90±0.14 | 1.20±0.10 |
| $-U^*_g$ [$10_{33}$ J] | 3300±300 | 30±3 |

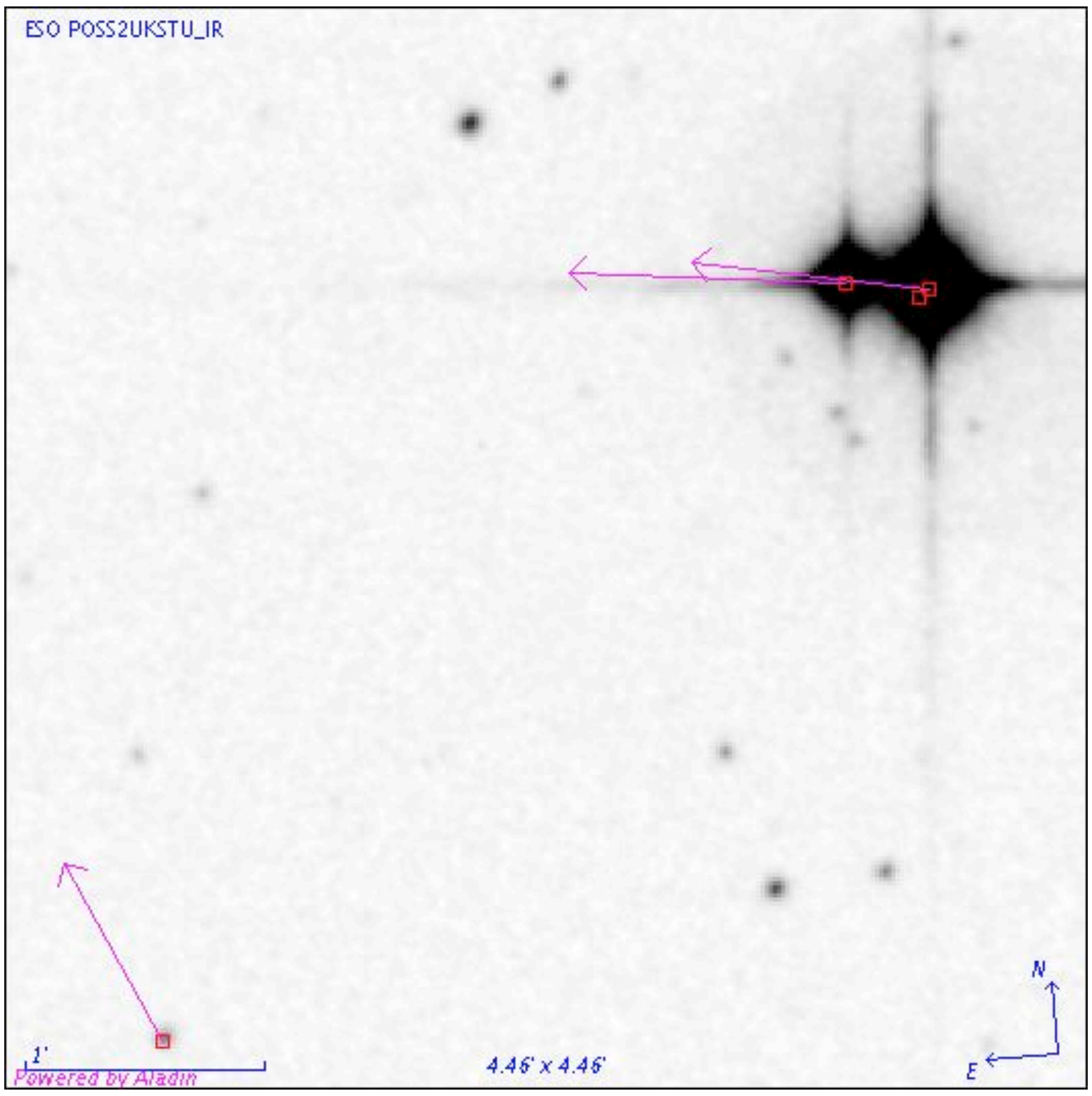

FIG. 1

Colour-inverted UKST $I_N$-band image provided by *The Digitized Sky Survey* from the European Southern Observatory at Garching constructed with *Aladin* showing the components A (HD 212168; top right corner, western star), B (CD−75 1242; top right corner, eastern star) and C (DENIS J2226443−750342, Koenigstuhl 5 C; bottom left corner). The long (pink) arrows indicate the proper motions as tabulated by *Simbad*. Note the incorrect proper motion of component C as measured by Schmidt *et al.*[41]. Field of view and orientation are also indicated on the image.